\documentclass[conference]{IEEEtran}
\IEEEoverridecommandlockouts
\usepackage{cite}
\usepackage{amsmath,amssymb,amsfonts}
\usepackage{algorithmic}
\usepackage{graphicx}
\usepackage{subcaption}
\usepackage{url}
\usepackage{pgf}
\usepackage{textcomp}
\usepackage{xcolor}
\usepackage{tikz}
\usepackage{balance}

\usepackage{fancyhdr}
\def\BibTeX{{\rm B\kern-.05em{\sc i\kern-.025em b}\kern-.08em
    T\kern-.1667em\lower.7ex\hbox{E}\kern-.125emX}}
\begin{document}
\newcommand\note[2]{\color{#1}\bf #2}
\newcommand\pb[1]{{\note{red}{paolo: #1}}}
\newcommand\af[1]{{\note{blue}{angelo: #1}}}
\newcommand\ag[1]{{\note{green}{andrea: #1}}}

\newcommand*\numcircledmod[1]{\raisebox{.5pt}{\textcircled{\raisebox{-.9pt} {#1}}}}

\newcommand\omnetpp{{OMNeT++}}
\newcommand\wifi{{Wi-Fi}}

\IEEEspecialpapernotice{%
    \begin{minipage}{\textwidth}
        \centering
        \textbf{Status: Work-in-Progress}\\[0.5em]
        This paper represents an ongoing project and includes preliminary ideas and results.\\
        It is not currently under review or accepted for publication.
    \end{minipage}
}
\title{Controlling Communications Quality in V2V Platooning: a TSN-like Slot-Based Scheduler Approach}

\author{\IEEEauthorblockN{Angelo Feraudo, Andrea Garbugli, Paolo Bellavista}
\IEEEauthorblockA{\textit{Department of Computer Science and Engineering} \\
\textit{University of Bologna}\\
Bologna, Italy \\
\{name.surname\}@unibo.it}
}

\maketitle

\begin{abstract}
Connected vehicles, facilitated by Vehicle-to-Vehicle (V2V) communications, play a key role in enhancing road safety and traffic efficiency. However, V2V communications primarily rely on wireless protocols, such as Wi-Fi, that require additional collision avoidance mechanisms to better ensure bounded latency and reliability in critical scenarios. In this paper, we introduce a novel approach to address the challenge of message collision in V2V platooning through a slotted-based solution inspired by Time-Sensitive Networking (TSN), which is gaining momentum for in-vehicle networks. To this end, we present a controller, named TSNCtl, operating at the application level of the vehicular communications stack. TSNCtl employs a finite state machine (FSM) to manage platoon formation and slot-based scheduling for message dissemination. The reported evaluation results, based on the \omnetpp{} simulation framework and INET library, demonstrate the effectiveness of TSNCtl in reducing packet collisions across various scenarios. Specifically, our experiments reveal a significant reduction in packet collisions compared to the CSMA-CA baseline used in traditional \wifi-based protocols (e.g., IEEE 802.11p): for instance, with slot lengths of 2 ms, our solution achieves an average collision rate under 1\%, compared to up to 50\% for the baseline case.
\end{abstract}


\section{Introduction}

In the current landscape of transportation, vehicles have experienced a significant transformation, evolving into interconnected entities equipped with a range of sensors, driver assistance, and safety-related systems. This evolution has driven car manufacturers, standard institutes, academia, and government agencies to enable vehicles and network infrastructure with new communications modes, e.g., Vehicle-to-Vehicle (V2V), Vehicle-to-Infrastructure (V2I), and Vehicle-to-Pedestrian (V2P). The necessity for vehicles to communicate arises from the need to facilitate numerous advanced applications leveraging vehicular technologies to enhance road safety, to provide advanced infotainment opportunities, to optimize traffic, to enable autonomous driving, and to support advanced manufacturer services~\cite{baldessari2007car, etsi2009etsi, car2car}. For instance, vehicles can leverage V2V communications to form dynamic platoons, i.e., a cooperative group of vehicles united in a common purpose, such as traffic flow optimization and infotainment services~\cite{jia2015survey,zeadally2020tutorial}.

Central to enabling these networks are the underlying communication protocols, typically based nowadays on \wifi{} and cellular technologies. The initial effort toward enabling connectivity between vehicles was the development of the IEEE 802.11p~\cite{5514475} protocol, standardized in the EU with the Intelligent Transport Systems (ITS) G5 (ITS-G5) framework~\cite{etsi302} and in the U.S. with Wireless Access in Vehicular Environment (WAVE) framework~\cite{ieee2016ieee}. In addition, the 3rd Generation Partnership Project (3GPP) leverages the existing cellular network infrastructure to offer a unified solution for V2X communications~\cite{release14,release16}. These protocols play a central role in enabling seamless data exchange, by enabling vehicles to transmit also safety-critical information, such as position, speed, and trajectory, to nearby vehicles on time.


Despite the advancements in V2V communication protocols, future applications like remote and autonomous driving demand meticulous resource coordination and control among vehicles~\cite{zang2019,jeon2018,maruko2018}. These scenarios heavily rely on messages for cooperative awareness and safety applications, which are broadcast without acknowledgments~\cite{cam, den}. The loss of such messages could disrupt the correctness of some critical functionality, such as for autonomous driving, by non-negligibly increasing the risk of accidents. Moreover, the transmission of cooperative awareness messages occurs at a given frequency, thus exacerbating the probability of data loss and leading to repeated interference between nodes~\cite{jeon2018}. Consequently, mitigating interference and data loss calls for the development of effective and efficient collision avoidance solutions, specifically tailored to vehicular communication scenarios.


We claim that the next generation of solutions for V2V communications hinges on the integration of Time-Sensitive Networking (TSN), a technology that offers precise timing synchronization and deterministic communication capabilities~\cite{nasrallah2018ultra}. This integration extends beyond V2V communication, encompassing inner-vehicle communication networks as well as broader V2X communication frameworks~\cite{satka2023comprehensive}. By leveraging TSN-based slotted protocols and exploiting time synchronization mechanisms, it becomes possible to orchestrate coordinated message transmission schedules, thereby mitigating the risk of message collisions and enhancing the overall communication reliability.

In this paper, we propose an original TSN-like controller, named TSNCtl, that aims at facilitating data delivery within car platooning based on V2V communications. For enhanced flexibility and dynamic deployment, it operates at the application layer and employs a finite state machine (FSM) to help in platoon formation and message dissemination for intra-platoon communications. The platoon creation/joining procedure starts upon receiving ITS service-related messages or detecting to be in the neighborhood of an existing platoon. Once a vehicle becomes part of a platoon, TSNCtl intercepts each message generated by the ITS service running on the vehicle, determining the priority queue for its storage. Next, it disseminates the messages based on its allocated time slots and the priority of each message.

To validate the effectiveness of our proposed approach, we have implemented our TSNCtl solution on the OMNeT++ simulation library and employed INET 4.5 to replicate V2V scenarios in accordance with the IEEE 802.11p standard. Next, we conducted a series of experiments aimed at evaluating performance metrics, by particularly focusing on packet collisions resulting from vehicle communications within a platoon. Our experiments clearly indicate that our solution outperforms traditional collision avoidance methods utilized in IEEE 802.11 protocols, such as CSMA-CA. Specifically, when considering slot lengths exceeding 1 ms, our proposal exhibits an average collision rate of less than 1\%, in contrast to up to 50\% collision rate observed with more traditional approaches.

The remainder of the paper is organized as follows. In Section \ref{sec:background}, we present the needed background about TSN-based in-vehicle communications, vehicular networks, and related simulation tools. Section \ref{sec:related} analyses the existing solutions for collision avoidance in V2V communications. Next, Section \ref{sec:sysarch} presents the details of our original TSNCtl controller. We then evaluate our proposal in section \ref{sec:evaluation}, by focusing on packet collisions in intra-platoon communications. Finally, we conclude with a discussion on the current limitations of our approach and on the main extensions that are the subject of our ongoing research work.
\begin{figure}[t]
    \centering
    \includegraphics[width=\linewidth]{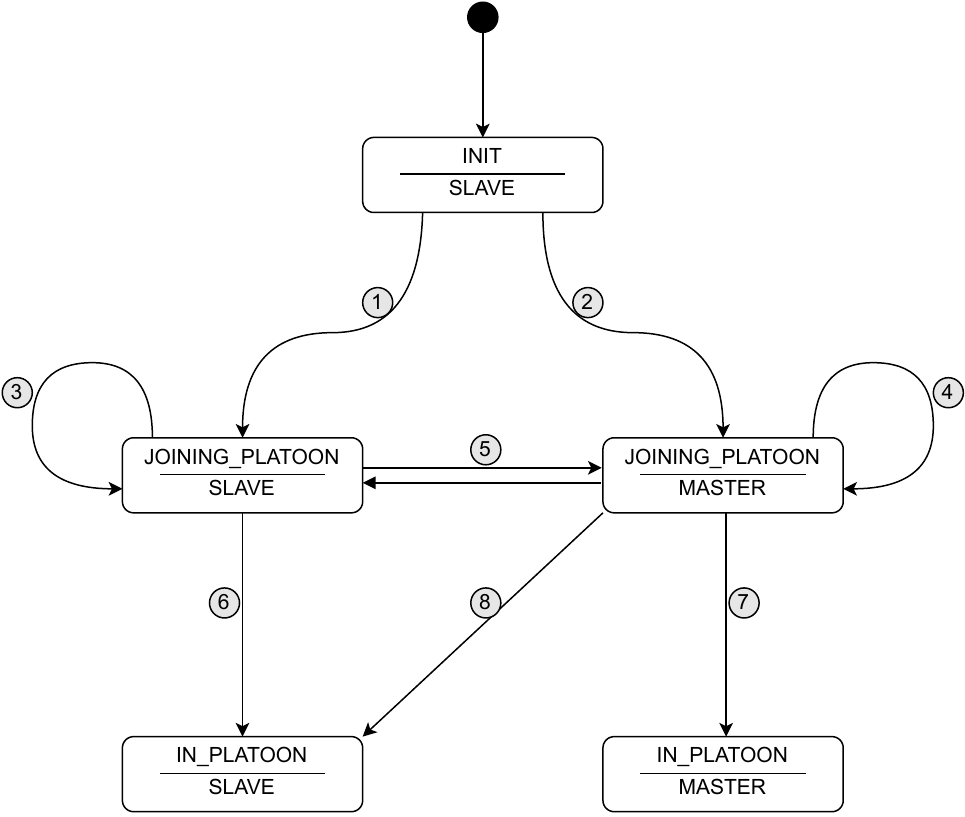}
    \caption{State Diagram Message Controller}
    \label{fig:ctlState}
\end{figure}

\section{Background}\label{sec:background}

This section provides a brief overview of the protocols used in V2X communications, with a specific emphasis on the Wi-Fi technologies that underlie V2V communications. It also introduces the suite of TSN protocols used for in-vehicle communications, along with an exploration of the simulation tools used to model complex vehicular networks, i.e. the \omnetpp{} framework, which allows for simulation of both in-vehicle and V2V communication protocols.

\subsection{TSN for in-vehicle communication}\label{sub:tsninv}

Time-Sensitive Networking (TSN) comprises a suite of standards designed to make Ethernet networks deterministic, particularly to accommodate real-time traffic flows. In the field of automotive Ethernet communications, TSN outlines a specialized profile known as IEEE 802.1DG, tailored to meet the rigorous requirements of in-vehicle networks (IVNs) in terms of high reliability and low latency~\cite{peng2023}.

The 802.1DG profile was developed to provide a set of mechanisms that ensure the timely delivery of crucial traffic, particularly control and security messages while supporting best-effort traffic.

In applications using TSN protocols, a crucial component is to establish a reliable time synchronization mechanism that ensures that all communication entities operate based on a unified time reference. This synchronization is facilitated within TSNs by the IEEE 802.1AS protocol, an extension of the IEEE 1588 Precision Time Protocol (PTP) known as generic PTP (gPTP)~\cite{8021AS}. Central to the PTP specification is the concept of a grandmaster clock, typically a highly accurate and stable source such as a Global Positioning System (GPS) receiver. This grandmaster clock serves as the central provider of reference time for all devices in the network. In the context of gPTP, devices in the network are designated as Clock Master (CM) and Clock Slave (CS), each of which participates in the exchange of synchronization messages. The CM disseminates time information to the connected CSs via multicast communication. Then, each CS, known as a gPTP instance, adjusts its synchronized time taking into account the message propagation delay along the gPTP communication path from the master to the instance. Once synchronization is achieved among all devices, a cohesive and time-aware network emerges, forming the so-called gPTP domain.

For Quality of Service (QoS), the TSN working group defines several traffic shaping techniques~\cite{nasrallah2018ultra}. In particular, IEEE 802.1Qbv introduces a time shaper known as the Time-Aware Shaper (TAS). The TAS orchestrates the scheduling of network frames belonging to different types of time-critical flows~\cite{garbugli2021end}. The standard outlines time-aware communication windows, each associated with a specific transmission queue. These windows are further segregated into cyclically repeating time slots, allowing frames to be buffered until the next associated time slot is opened. This segregation ensures that assured traffic maintains low latency and minimal jitter while avoiding interference from other traffic streams. The definition of windows and slots is facilitated by a Gate Control List (GCL), which identifies the time instants at which queues are open for frame transmission. As a result, TAS can meet the requirements of ultra-low latency and reliability, provided that all time windows are synchronized, which makes its combination with PTP necessary.

\subsection{V2X Communications and Protocols}
In Vehicular Ad-Hoc Networks (VANET) vehicles exploit On Board Unit (OBU) network interface to propagate messages with other vehicles within their proximity. These networks define various communication modes aimed at automating message dissemination~\cite{zeadally2020tutorial}: Vehicle-to-Vehicle (V2V), Vehicle-to-Infrastructure (V2I), Vehicle-to-Pedestrian (V2P), and Vehicle-to-Everything (V2X). Communications modes are enabled by both WiFi-based technologies, such as IEEE 802.11p/bd~\cite{5514475}, and Cellular-based technologies like LTE and NR C-V2X~\cite{release14,release16}.  

This proposal primarily focuses on WiFi-based technologies and V2V mode, considering the exchange of messages for cooperative awareness \cite{cam} and safety-related applications \cite{den}. Thus, Cellular-based standards and other communication modes pertaining to the vehicular environment are beyond the scope of this paper.

IEEE 802.11p is specifically tailored to support vehicular communication requirements. It introduces enhancements to accommodate features such as vehicle velocities up to 200 km/h and communication ranges up to 100m. Operating at the physical and Medium Access Control (MAC) layers, IEEE 802.11p enables connectivity between vehicles and infrastructure. At the physical layer, it employs Orthogonal Frequency Division Multiplexing (OFDM), while at the MAC layer, it utilizes the Enhanced Distributed Channel Access (EDCA) method, incorporating carrier sense multiple access with collision avoidance (CSMA/CA), without exponential back-off and message acknowledgment. 

In 2019, the Internet Task Force embarked on the development of a new standard for V2X communication, known as IEEE 802.11bd. This standard aims to supersede IEEE 802.11p by offering twice the performance in terms of throughput, latency, reliability, and communication range \cite{yacheur2020}. IEEE 802.11bd also introduces support for message retransmissions and ensures backward compatibility with its predecessor.

\subsection{\omnetpp~and INET}
\omnetpp~\cite{omnetwebsite} is a well known discrete-event simulation framework capable of modeling various network types and communications. It relies on the concept of modules, fundamental elements that facilitate message exchange through connections and gates. Their behavior is defined in C++, while descriptions, including gates, connections, and parameters, are expressed in Network Description language (NED). Initialization files (INI) are used to specify parameter values for model initialization, allowing for multiple values or intervals to be defined for parameters. With its provision of basic simulation functionalities, \omnetpp{} allows users to concentrate on creating their simulation models.

The INET~\cite{inet} framework is a library built on top of \omnetpp. It furnishes models for a myriad of network components, including communication protocols, network nodes, and connections. Furthermore, it features Internet stack models and wired and wireless link layer protocols. This enables users to instantiate and interconnect protocol layers, facilitating the creation of custom hosts tailored to their specific requirements. This modular framework enables rapid configuration of complex models, thereby enhancing efficiency in simulation setup.

\section{Related Work}\label{sec:related}  
Platooning plays a crucial role in advanced vehicular applications, enabling vehicles to cooperate towards a common goal. However, vehicular networking performance, such as packet loss and transmission delay, can significantly affect platoon dynamics~\cite{jia2015survey}. 

Addressing these challenges, Shao \textit{et al.}~\cite{shao2015performance} introduced a MAC-layer protocol able to deal with the intermittent connectivity of VANETs. This protocol relies on multichannel reservation and priority schemes that enhance throughput and packet delivery ratio. Similarly, in~\cite{jonsson2013increased} the authors proposed a polling-based MAC-layer protocol coupled with a customized transport layer retransmission scheme that reduces message error rates within intra-platoon communications. 

Other existing proposals operate under the assumption that a designated platoon leader is tasked with allocating time slots, similar to Time Division Multiple Access (TDMA), to other platoon members, thus defining specific time windows for their communication activities~\cite{fernandes2012platooning,segata2014towards,park2012collision}. For instance, in~\cite{fernandes2012platooning} Fernandes \textit{et al.} provided a system that divides the control channel of the IEEE 802.11p into slots to ensure stable intra-platoon communications. In the same directions, Segata \textit{et al.} proposed a beacon dissemination strategy based on TDMA in~\cite{segata2014towards}, where the platoon leader takes the lead in beacon transmission, followed by other platoon members. 


From a collision avoidance perspective, Zang \textit{et al.}~\cite{zang2019} proposed a collision detection approach, where transmitting nodes employ full-duplex channel sensing to detect simultaneous transmissions. In cellular-based environments (C-V2X), researchers are exploring methods to enhance scheduling algorithms, aiming to reduce or detect packet collisions in scenarios where vehicles are out of base stations coverage (V2V)~\cite{maruko2018,jeon2018,cao2022resource}. 

Despite the research efforts to enhance the reliability of vehicular network communications, there remains a need for more building solutions that are technology-agnostic, regardless of whether they are based on \wifi or cellular technology. Moreover, it is crucial to consider larger-scale scenarios without solely relying on mathematical models.

In our proposal, we have developed a TSN-like controller at the application layer, which operates independently of the specific wireless technology utilized in the vehicle. This controller is designed to enhance the packet delivery ratio in intra-platoon V2V communications by establishing platoons, wherein each vehicle has a dedicated slot for message dissemination. The adoption of a TSN-based scheme aligns with the emerging trend in both car manufacturing and academia, where there is a growing interest in utilizing this technology in in-vehicle communications systems. 

\begin{figure}[t]
    \centering
    \includegraphics[width=0.5\linewidth]{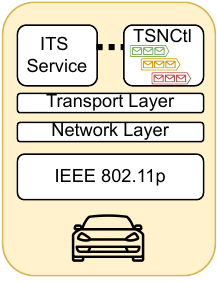}
    \caption{Integration of Message and Platoon Controller in V2X Stack}
    \label{fig:stack}
\end{figure}

\section{System Architecture}\label{sec:sysarch}
\subsection{General Overview}
As mentioned in Section \ref{sub:tsninv}, the future of in-vehicle communications is increasingly focused on exploiting TSN to ensure low-latency, reliability, and determinism for critical vehicular services \cite{peng2023}.  Our proposal relies on the concept that TSN-enabled vehicles could leverage Radio and GPS technologies \cite{Baniabdelghany2020, Alemdar2021, garg2018, hasan2018} to synchronize themselves, facilitating synchronized and time-sensitive communications over wireless connectivity. Thus, in our proposal, each vehicle is equipped with a TSN-like controller responsible for orchestrating the dissemination of messages in V2V communications (in this work we do not address at all vehicle clock synchronization).

In Figure \ref{fig:stack}, we observe the integration of the proposed controller, \textit{TSNCtl}, at the application layer of the V2X stack. TSNCtl operates an FSM that helps vehicles in platoon formation and engages in V2V communications through slot-based scheduling. The platoon formation process regards determining which vehicles partake in the communications network. In our scenario, this relies on \textit{TSNCtl} FSM, which initiates platoon creation or joining procedures upon receiving ITS service-related messages or upon detecting proximity to an existing platoon. During this procedure, a platoon \textit{master} is elected among the vehicles, and tasked with allocating slots to each member of the platoon. These slots are utilized by vehicles to communicate at designated times, minimizing the risk of collisions during communications with other vehicles within the platoon. It should be noted that a vehicle may request multiple slot allocations if it hosts multiple ITS or safety-related services.

Once the platoon has been established, \textit{TSNCtl} intercepts messages generated by the ITS Service running on the vehicle and, based on the slot allocated for that vehicle, determines the optimal timing for their dissemination across the network. To manage this process effectively, \textit{TSNCtl} maintains a series of queues, each assigned with a distinct priority level, wherein these messages are stored before transmission. Depending on the nature of the message, such as safety or non-safety related, \textit{TSNCtl} determines the appropriate queue for its placement and subsequent dispatching.

\begin{figure}[t]
    \centering
    \includegraphics[width=\linewidth]{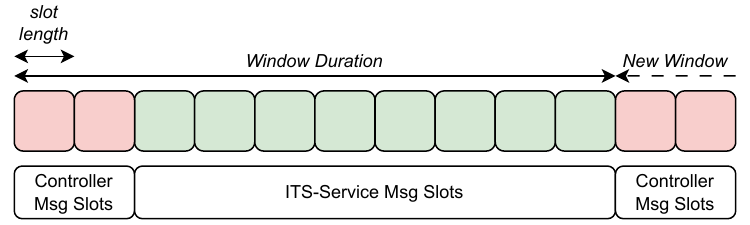}
    \caption{Slot-based Scheduling}
    \label{fig:slots}
\end{figure}
\begin{figure*}[t]
    \begin{subfigure}[b]{0.5\textwidth}
       \centering
        \resizebox{\linewidth}{!}{\input{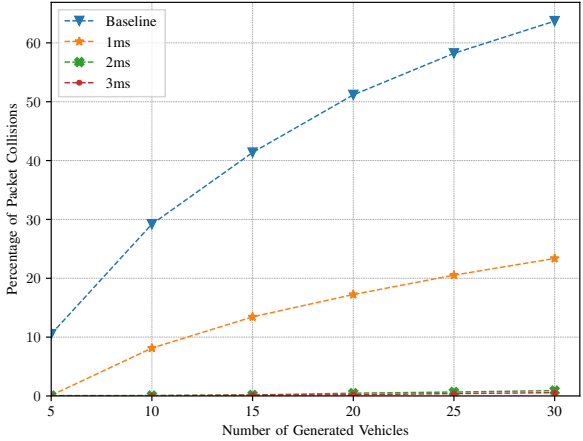}}
        \caption{Packet Collisions with a $100 \mu s$ Vehicle Spawning Frequency}
        \label{fig:100micros}
    \end{subfigure}
    \begin{subfigure}[b]{0.5\textwidth}
        \centering
        \resizebox{\linewidth}{!}{\input{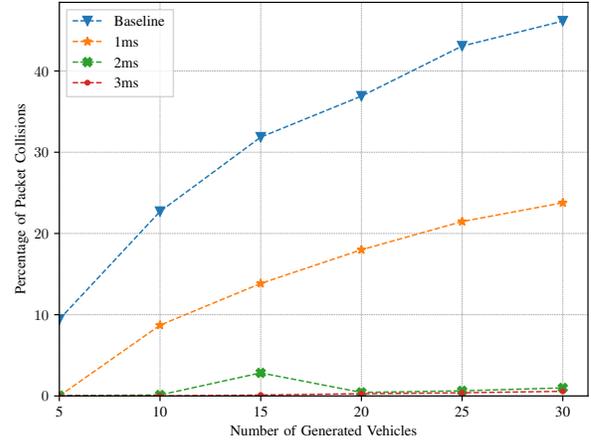}}
        \caption{Packet Collisions with a $1ms$ Vehicle Spawning Frequency}
        \label{fig:1ms}
    \end{subfigure}
    \caption{Collisions in Packet Transmission with Different Platoon Size}
    \label{fig:controllerPerf}
\end{figure*}

\subsection{TSNCtl: Platoon Formation and Slot Assignment}

The \textit{TSNCtl} FSM is used to model the behaviors that vehicles exhibit during platoon formation. Figure \ref{fig:ctlState} illustrates the states that the FSM can assume. Each state is defined by a combination of node status, including \textit{init}, \textit{joining\_platoon} and \textit{in\_platoon}, and a node role,\textit{slave} and \textit{master}, defining vehicle responsibilities within the platoon.

During the initialization phase, the vehicle is in the \textit{init} state, wherein parameters such as communication window duration and slot length are established. The communication window duration sets the interval for vehicle coordination within the platoon. Meanwhile, the slot length determines the slot duration, which is used to select when a vehicle can transmit its messages regulating the timing of data dissemination.

In the \textit{init} state, the internal \textit{TSNCtl} component remains idle until the specified window duration elapses. Referring to Figure \ref{fig:slots}, if the communication window duration is set to 100 ms, controller-based messages can only be exchanged at intervals that are multiples of this time interval (window). The number of available slots within a window is determined by the slot length; for example, with a 100 ms window and a 10 ms slot length, the total available slots amount to 10.

After the window elapses, the first two slots are designated for controlling operations, specifically aimed at forming or joining a platoon (red slots in Fig. \ref{fig:slots}). In the initial time slot, controller messages are transmitted, conveying node-related information alongside the timestamp of message generation. It is worth noting that, in this phase, nodes compete to become masters of the platoon. Thus, controller messages are randomly generated within the interval defined by the slot length, to prevent collisions during this slot.

During the second time slot, \textit{TSNCtl} examines the content of the messages received in the previous time slot and changes its status in \textit{joining\_platoon}. As illustrated in Figure \ref{fig:ctlState}, in steps \numcircledmod{1} and \numcircledmod{2} the node's role in this state depends on the timestamps gathered thus far: the node with the earliest timestamp, representing the first generation of the controller message, assumes the role of \textit{master}, while the others become \textit{slaves}.  Within this slot, the \textit{master} node dispatches a message containing details about slot allocations. These allocations can be determined using various policies, potentially influenced by factors such as the number of slots requested by a node and the type of node (e.g., car, ambulance, etc.).

However, the \textit{master} node requires at least one neighboring node to establish a platoon. Thus, if no node is detected nearby, the \textit{master} node maintains the current status and restarts the procedure to form the platoon (step \numcircledmod{4}). A \textit{master} node persisting in this state, may become \textit{slave} at the next iteration of the controller-related messages (step \numcircledmod{5}). Conversely, in the presence of other nodes, its state transits in \textit{in\_platoon} (step \numcircledmod{7}).

On the other side, \textit{slave} nodes await the slot allocation defined by the \textit{master} and prepare to transmit their data during their allocated time slot. Upon receiving the internal trigger corresponding to their slot, \textit{slave} nodes confirm their role and transition to the \textit{in\_platoon} state (step \numcircledmod{6}). However, those \textit{slave} nodes that fail to receive a message from the \textit{master} remain in this state, step \textcircled{3} in Figure \ref{fig:ctlState}, and restart the platoon joining procedure during the next TSNCtl slots.

In the \textit{in\_platoon} state, TSNCtls with a \textit{slave} role transmit messages solely during their designated slots, while \textit{master} nodes also utilize the second slot reserved for controller communication. The latter allows the integration of new nodes into the platoon.

Following platoon formation, each message generated by the ITS service is categorized based on its priority and inserted into one of the available queues within the controller. The TSNCtl then disseminates these messages according to the allocated slots and message priorities.


\section{Evaluation Results} \label{sec:evaluation}

To demonstrate the usefulness and efficiency of our proposal, we developed the TSNCtl component outlined in Section \ref{sec:sysarch} within the \omnetpp~ simulation library. We utilized INET 4.5 to simulate ad-hoc communications and specifically V2V scenarios, adhering to the IEEE-802.11p standard.
We implemented two \omnetpp~ simple modules: a mock service, sending messages to the controller (Fig. \ref{fig:stack}), and a mock application, sending messages directly on the socket. Both modules generate UDP packets with a regular interval of 100 ms, by following the ETSI standard for Cooperative Awareness\cite{cam}. Packet sizes relies on the analysis conducted by the CAR2CAR Communication Consortium \cite{martinez_survey_2018} in real-world scenarios. Furthermore, we introduced a \textit{spawner} module that generates vehicles within a specified area and frequency. For our testing, we considered an area for platoon formation that corresponds to the V2V communication coverage range (between 100 to 300 meters) and different frequencies for vehicle spawning (1 ms and 100 $\mu s$).

We have conducted several experiments to assess the viability of our proposal. Initially, we explored the behavior of our model by varying the number of vehicles within a platoon and their entry frequency. Secondly, we have tested the resilience of our solution when changing packet size. These experiments were performed on a Linux VM running \omnetpp~having 16 CPUs and 32 GB of RAM. Each experiment was repeated five times to ensure statistical reliability. It should be noted that, at this stage, the experiments do not take into account signal attenuation due to obstacles like objects and foliage on the road. This approach also allowed us to correctly identify those packets that collided within the simulated scenario.

\subsection{Assessing Packet Collisions by Varying Platoon Sizes}\label{sub:pcplatsize}
The first set of experiments involves the investigation of the performance of our proposed solution (Section \ref{sec:sysarch}) when varying the dimension of the platoon size. These experiments consider 800 bytes as max packet size \cite{martinez_survey_2018, molina2020}. Furthermore, the window duration is set to 100 $ms$ and three slot lengths are analyzed: namely 1, 2, and 3 $ms$. The graphs showed in Figure \ref{fig:controllerPerf} compare the percentage of packet collisions in intra-platoon communications between scenarios without \textit{TSNCtl} (\textit{Baseline}) and those using it.

Figure \ref{fig:100micros} shows the percentage of packet collisions obtained when vehicles join the platoon with a frequency of $100\mu s$, while Figure \ref{fig:1ms} with $100ms$ generation frequency. The \textit{spawner} module generates vehicles within the simulation area of 100 meters at the specified frequency. Upon generation, each vehicle starts broadcasting messages every 100$ms$. In both scenarios, it is evident that the proposed \textit{TSNCtl} correctly avoids packet collisions and helps in platoon formation. However, when the slot length is below 2$ms$, the controller performance is significantly reduced, from an average of 0.40\% and 0.23\% packet collisions for the 2$ms$ and 3$ms$ cases, to an average of 13.8\% with slot length of 1$ms$ (Figure \ref{fig:100micros}). The indicator reaches around 20\% of collisions when the platoon size is above 20 vehicles. Similarly, widening the gap between vehicle generation (Figure \ref{fig:1ms}) results in collision rates approaching 25\% for platoon size over 20 vehicles.

Hence, slot lengths shorter than 2$ms$ fail to allow messages to propagate in time from the source to all destinations within the platoon, leading to packet interference. Furthermore, we have experiences that this issue gets worse when considering signal attenuation caused by obstacles and foliage, although this aspect is not specifically addressed yet in this paper. Despite the poor performance with small slot length, anyway our controller is still able to synchronize platoon vehicles, by outperforming more traditional scenarios. As illustrated in both Figure \ref{fig:100micros} and \ref{fig:1ms}, the \textit{Baseline} performance yields a minimum of 10\% of collisions during platoon formation, while our solution does not exceed 1\% when choosing the appropriate slot length.

\subsection{Assessing Packet Collisions by Varying Packet Sizes}\label{sub:pcpacksize}
\begin{figure}[t]
    \centering
    \resizebox{\linewidth}{!}{\input{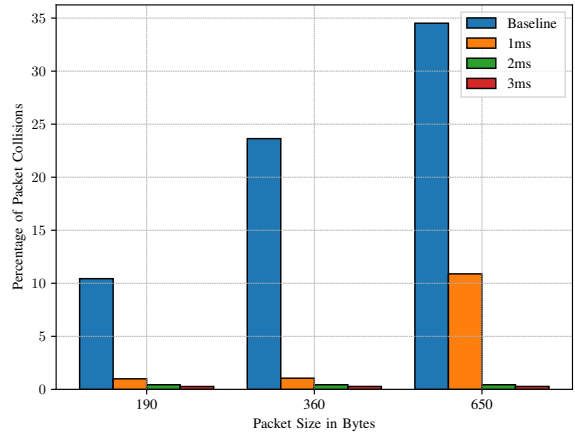}}
    \caption{Collisions in Packet Transmission with Different Packet Size}
    \label{fig:packetsize}
\end{figure}

In these experiments, we examine a platoon consisting of 20 vehicles generated within a 100-meter area at intervals of 1$ms$. Window duration is fixed to 100 $ms$, while different slots lengths are analyzed. Figure \ref{fig:packetsize} illustrates the percentage of collisions observed while varying packet dimensions in this scenario. The chosen packet dimensions for this analysis are derived from the data provided  \cite{martinez_survey_2018}.

The overall trend depicted the \textit{Baseline} bar, is an increase in collisions as packet dimensions grow. It shows an approximate 12\% increase in collisions with each incremental change in packet dimension. Conversely, the controller shows a consistent collisions rate below 1\% when using slot length above 1$ms$. However, as also occurred in the previous analysis (Section \ref{sub:pcplatsize}), using short slots fail to adequately avoid collisions. As the picture demonstrates, for the largest packet size considered (650 bytes) the packet collisions exceed the 10\% of the total message sent.

\section{Discussion and Future works}
Efficiently achieving intra-platoon message dissemination continues to be a challenging issue. To this purpose, we propose a TSN-like controller that leverages the capabilities of in-vehicle networks based on the TSN~\cite{peng2023} approach. This direction is motivated by the growing enthusiasm of academia and industry for the use of TSN in wireless environments for message propagation~\cite{thi2022ieee}, in particular thanks to the introduction of the latest IEEE 802.11 standard, i.e. IEEE 802.11be, also called \wifi{} 7~\cite{adame2021time}, which aims to fully integrate TSN functionality to support low-latency and ultra-reliable communications. 

\noindent \textbf{5G Integration.} Our approach not only opens up new challenges and directions for research but also holds the potential to profoundly change the landscape of message dissemination within VANETs. A TSN-enabled V2V environment can boost both packet delivery ratios, as demonstrated in this paper, and network performance in large-scale VANET deployments. Furthermore, the integration of the cellular and TSN domains is considered a promising solution for supporting time-sensitive applications and low-latency communications, especially in the case of 5G infrastructures~\cite{satka2023comprehensive}. This will bring several benefits also in V2V communications based on cellular technologies.

\noindent \textbf{Platoon Formation.} Looking ahead, our future research will involve extending our solution to realistic scenarios that consider signal attenuation generated by obstacles and vehicle trajectories, while also creating effective and smart strategies to handle errors in platoon formation. Moreover, vehicles that reside at the border of two different platoons must be handled accordingly, and can also be exploited as relays to forward messages between platoons, thus extending the communication range and facilitating seamless integration between adjacent platoons.

\noindent \textbf{Synchronization.} Our current implementation of the proposed solution is based on a common GPS-based time reference, i.e., all vehicles participating in the platoon are already synchronized before the platoon is formed. In a more complex scenario where not all participants are constrained to host GPS equipment, time synchronization opens up additional technical challenges. Related to that, the evolution of our solution will include a first phase in which the vehicles try to synchronize their clocks through the use of a wireless implementation of the PTP protocol; only once a synchronized time domain has been built, the actual communication can start.

\noindent \textbf{Messages Prioritization.} In complex communication scenarios, vehicles may want to exchange messages with different priorities. In this case, the design and implementation of different queues for differentiated message priority, especially in delay-sensitive scenarios, could take advantage of the TSN-based approach adopted by our solution. Distinguishing between safety-critical and non-safety-related messages allows for more efficient resource allocation and ensures timely delivery of critical information, thus improving the overall system reliability. In addition, we envision integrating our proposal within a cellular-based domain, particularly in out-of-coverage scenarios where vehicles may exploit the new V2X Mode 2~\cite{v2x3gpp} radio. The integration of distinct priority-based queues will also facilitate the reliable and rapid dissemination of safety-critical messages between different platoons.

\section{Conclusions}

This study introduces an innovative approach to improve V2V communications through a TSN-like slot-based scheduling solution, as a first step towards deterministic communication within vehicular networks. Our TSNCtl effectively orchestrates message dissemination, by mitigating packet collisions and by ensuring reliable data exchange between vehicles. The reported performance results, measured on top of the \omnetpp{} simulation environment, highlight the effectiveness of our approach, in particular in scenarios with different platoon sizes and packet sizes. Going forward, our focus will shift to refining the TSNCtl component to adapt it to different traffic conditions and to incorporate priority-based message queuing; in addition, we will better explore the implications of different network performance parameters and will evaluate the feasibility of operating the controller without prior synchronization.

\balance

\bibliographystyle{IEEEtran}
\bibliography{references}

\end{document}